\documentclass[aps,prl,twocolumn,showpacs,superscriptaddress]{revtex4-1}
\usepackage{graphicx,hyperref}

\begin{document}

\title{Asymmetry-induced resistive switching in Ag-Ag$_{2}$S-Ag memristors enabling a simplified atomic-scale memory design}

\author{A. Gubicza}
\affiliation{Department of Physics, Budapest University of Technology and Economics} \affiliation{MTA-BME Condensed Matter Research Group, Budafoki ut 8, 1111
Budapest, Hungary}
\author{D. Zs. Manrique}
\affiliation{Physics Department, Lancaster University, Lancaster,
UK}
\author{L. P\'osa}
\affiliation{Department of Physics, Budapest University of Technology and Economics} \affiliation{MTA-BME Condensed Matter Research Group, Budafoki ut 8, 1111
Budapest, Hungary}
\author{C. J. Lambert}
\affiliation{Physics Department, Lancaster University, Lancaster, UK}
\author{G. Mih\'aly}
\author{M. Csontos}
\email{csontos@dept.phy.bme.hu}
\author{A. Halbritter}
\affiliation{Department of Physics, Budapest University of Technology and Economics} \affiliation{MTA-BME Condensed Matter Research Group, Budafoki ut 8, 1111
Budapest, Hungary}

\date{\today}

\begin{abstract}
Prevailing models of resistive switching arising from electrochemical formation of conducting filaments across solid state ionic conductors commonly attribute
the observed polarity of the voltage-biased switching to the sequence of the active and inert electrodes confining the resistive switching memory cell. Here we
demonstrate equivalent, stable switching behavior in metallic Ag-Ag$_{2}$S-Ag nanojunctions at room temperature. Our experimental results and numerical
simulations reveal that the polarity of the switchings is solely determined by the geometrical asymmetry of the electrode surfaces. By the lithographical
design of a proof of principle device we demonstrate the merits of simplified fabrication of atomic-scale, robust planar Ag$_{2}$S memory cells.
\end{abstract}


\maketitle

As ongoing miniaturization reaches the fundamental limitations of silicon-based complementary metal-oxide-semiconductor (CMOS) technology, the demand for
alternative material platforms delivering faster, smaller, yet highly integrable logical and memory units is increasing. Self-assembled nanostructures
exhibiting tunable electrical properties are primary candidates. Conducting nanofilaments formed or destroyed by reversible solid state electrochemical
reactions in ionic conducting media situated between metallic electrodes have demonstrated reproducible logical and non-volatile resistance switching random
access memory (ReRAM) operations
\cite{Yang2013,Borghetti2010,SolvingMazes,nature03190,nature06932,nmat2023,advmat2007,Wang2007,Jo2008,IEEE2010,FlashAndReRam,nmat2748,Pershin2011,Valov2011,Torrezan2011,Hasegawa2012}.
The resistance of such a two-terminal memristor \cite{1083337}, is altered above a threshold bias ($V_{\rm th}$) of a few hundred mV. Nonvolatile readout is
performed at $V<<V_{\rm th}$ \cite{Gubicza2015}.

Nanofilament formation in solid state electrolytes
\cite{nature03190,advmat2007,Chang2011,Ohno2011,wagenaar:014302,Strukov2008,Xu2010,Nayak2010,Masis2009,Masis2010,Masis2011,Nayak2011,Menzel2012,Valov2012,Yang2012,Hasegawa2012,Liu2012,Menzel2013,YangLu2013,Cheng2015}
are commonly attributed to oxidation, electric-field-driven ionic migration and reduction, involving a positively charged active electrode supplying the mobile
ions and a negatively charged inert electrode, where reduction can take place initializing the filament growth. At opposite polarity, the filament is
dissolved. While offering extremely large $R_{\rm OFF}/R_{\rm ON}$ switching ratios, devices operated in this regime can only perform at reduced switching
speeds due to their fundamental RC limitations. Once such a conducting channel is complete, smaller but orders of magnitude faster resistance changes can be
observed in the metallic regime \cite{Nardi2012,Larentis2012,Mickel2014,Torrezan2011,GeresdiAndreev,Gubicza2015,Gubicza2015temp}, which is also in the focus of
our present study.

The central question of our paper concerns the role of the inert electrode and consequently, the polarity of the set/reset transitions. Depending on the ionic
mobility and redox rates, nucleation and subsequent filament formation have been observed either at the inert or at the active electrode surfaces by in-situ
methods \cite{Xu2010,Liu2012,Yang2012,Yang2014}. After an initial nucleation phase further reduction takes place directly along the growing filament consisting
of the elemental metal of the active electrode. Thus, we anticipate that resistive switching must also occur when both electrodes are fabricated from the
active material. We propose that the polarity of the resistive switching is determined by the local inhomogeneity of the electric field, the latter reflecting
the geometrical asymmetry of the electrode surfaces with particular emphasis on the narrowest region of the filament.

Single-component metallic junctions utilizing silver \cite{C0NR00951B} and aluminium \cite{Schirm2013} break junctions exhibit reproducible resistive switching
at cryogenic temperatures, both relying on atomic rearrangements due to electromigration. Similar experiments were also carried out using nanofabricated gold
nanowires \cite{Johnson2010} at room temperature. While the operation speed of these devices were (instrumentally) limited to 100~kHz, Ag-Ag$_{2}$S-PtIr
memristive nanojunctions were successfully demonstrated to exhibit nanosecond switching times \cite{GeresdiAndreev}. However, the fabrication of such
ultrafast, Ag$_{2}$S based memory cells would be considerably simplified if a single metallic component could be utilized for both terminals of the devices.
Here we demonstrate stable resistive switchings in metallic Ag-Ag$_{2}$S-Ag nanojunctions created by an STM and alternatively, in a mechanically controllable
break junction (MCBJ) \cite{Muller1992} setup. The observed set/reset transitions exhibit a uniform polarity with respect to the initial asymmetry of the
junction geometry in agreement with our molecular dynamical simulations. The latter reveal the kinetics of filament growth and shrinkage under various boundary
conditions by taking into account electric-field-driven activated ionic migration in the Ag$_{2}$S matrix. Based on these findings a proof of concept all-Ag
on-chip device fabricated by standard electron beam lithography is testified to reliable memory operations.

\begin{figure}[t!]
\includegraphics[width=\columnwidth]{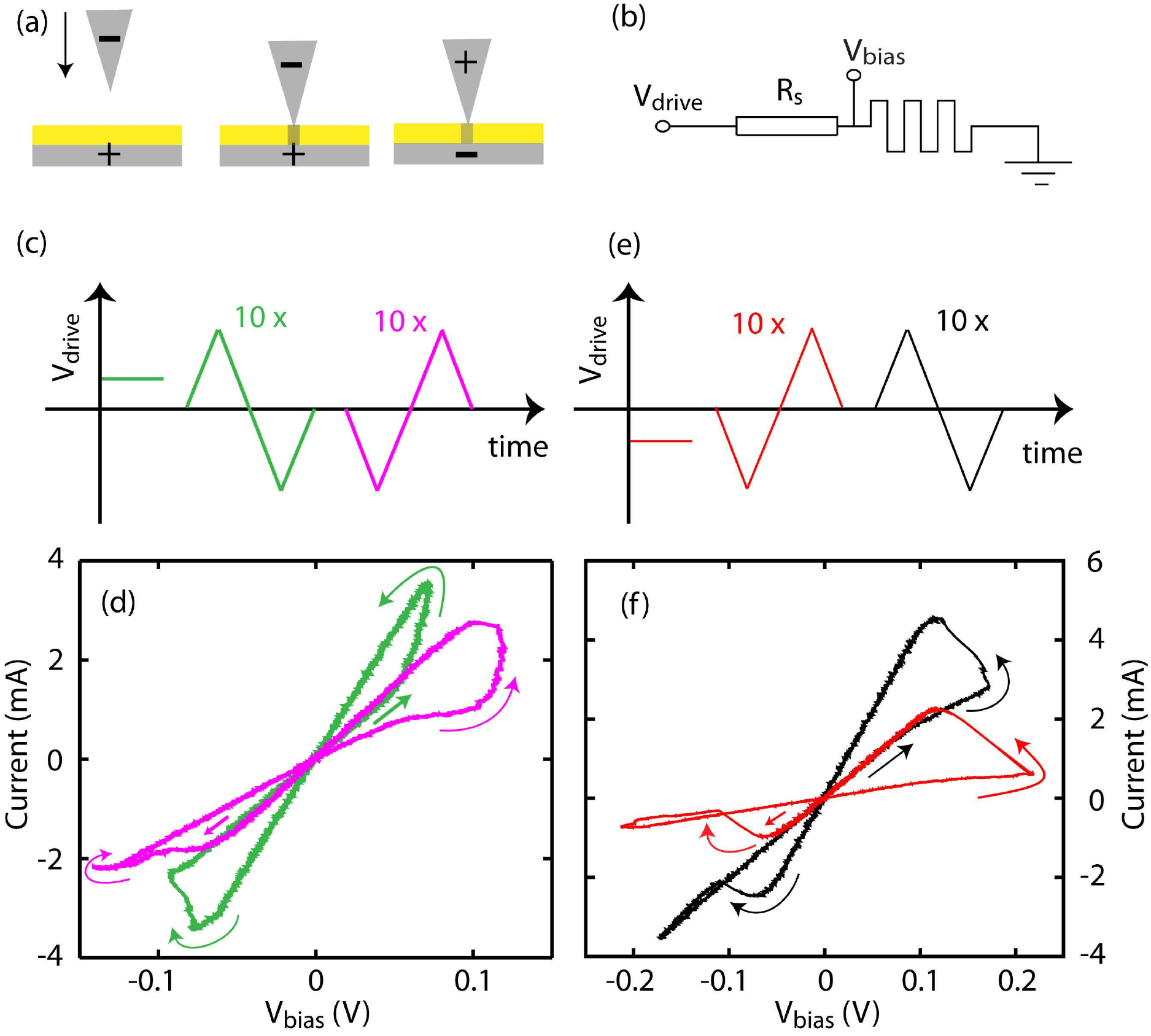}
\caption{(color online) (a) The scheme of the I-V measurements performed in the STM setup. A constant bias approach of the tip is followed by voltage sweeps of
alternating sign. By convention, a positive bias corresponds to a positive voltage applied to the Ag layer with respect to the STM tip. (b) The electrical
circuit diagram of the biasing setup. (c) A typical sequence of the triangular $V_{\rm drive}$ signals of 2.5~Hz. The junction is established in the presence
of a constant positive voltage of $V_{\rm drive}$ = 100~mV (green). Resistive switching behavior is investigated by a triangular $V_{\rm drive}$ starting with
a positive polarity (green) which is reversed after 10 periods (magenta). (d) The corresponding hysteretic I-V traces exhibiting a uniform switching direction
as indicated by the curved arrows. The straight arrows denote the initial configurations. (e) Approaching at a constant negative voltage of $V_{\rm drive}$ =
-100~mV (red) followed by a reversed sequence of $V_{\rm drive}$ with respect to (c) (red and black). (f) The corresponding I-V traces reveal identical
directions of the hysteresis loops to those in (d). The displayed device resistances are $R_{\rm OFF}$=38~$\Omega$, $R_{\rm ON}$=20~$\Omega$ (green), $R_{\rm
OFF}$=63~$\Omega$, $R_{\rm ON}$=34~$\Omega$ (magenta), $R_{\rm OFF}$=256~$\Omega$, $R_{\rm ON}$=51~$\Omega$ (red) and $R_{\rm OFF}$=49~$\Omega$, $R_{\rm
ON}$=23~$\Omega$ (black). $R_{\rm S}$=50~$\Omega$ and $V_{\rm drive}^{0}$=0.25~V.} \label{stm.fig}
\end{figure}

The first set of experiments utilized an 80~nm thick Ag layer deposited onto a Si substrate followed by a 5 minutes long sulfurisation performed at
60$^{\circ}$C resulting in a 30~nm thick stoichiometric Ag$_{2}$S cap layer on the planar Ag electrode \cite{GeresdiMRS2011}. Nanometer-scale junctions were
created between the Ag$_{2}$S surface and a mechanically sharpened 99.99\% pure Ag wire of 0.35~mm in diameter in STM geometry. Alternatively, Ag-Ag$_{2}$S-Ag
point contacts were also established by the controlled rupture of a 99.99\% pure Ag wire with a diameter of 0.125~mm in the vacuum chamber of an MCBJ setup
followed by a 20 minutes long in-situ sulfurisation carried out at 25$^{\circ}$C. During the acquisition of the current-voltage (I-V) characteristics the
$V_{\rm drive}$ low frequency triangular voltage output of a data acquisition card was acting on the memristive junction and on a variable series resistor
$R_{\rm S}$ as shown in Fig.~\ref{stm.fig}(b). The device's current was monitored by a current amplifier while the $V_{\rm bias}$ voltage drop on the junction
was determined numerically as $V_{\rm bias}=V_{\rm drive}-I\cdot R_{\rm S}$. All measurements were performed at room temperature.

Representative I-V traces obtained in the inherently asymmetric STM geometry are displayed in Fig.~\ref{stm.fig}. In order to investigate the influence of the
initial electroforming process on the direction of the observed resistive switching, measurements were performed on nanojunctions established with Ag tips,
which were either negatively or positively charged during approaching the Ag$_{2}$S thin films, as illustrated in Fig~\ref{stm.fig}(a). After forming a
metallic contact, a triangular $V_{\rm drive}$ voltage signal of the same initial polarity was applied to record the I-V traces over 10 periods. This was
followed by a reversed phase triangular $V_{\rm drive}$ of another 10 periods as indicated in Figs.~\ref{stm.fig}(c) and \ref{stm.fig}(e). The corresponding
I-V traces are exemplified in Figs.~\ref{stm.fig}(d) and \ref{stm.fig}(f). It is to be emphasized that all the four hysteresis loops share the same direction
of the resistive switchings, i.e., set (reset) transitions take place exclusively at positive (negative) biases on the Ag film, independently of the bias
polarity on approaching as well as of the initial field direction during the voltage sweeps. This provides a strong experimental evidence that the polarity of
the resistive transitions in the Ag$_{2}$S layer is solely determined by the inhomogeneity of the local electric field in the vicinity of the conducting
filament in accordance with the geometrical asymmetry of the surrounding Ag terminals. In order to gain a microscopic insight into the kinetics of field driven
filament evolution upon such biasing cycles and to understand the observed, robustly uniform polarity of the resulting resistive switchings we performed
atomic-scale numerical simulations taking all-Ag electrodes with various boundary conditions into account.

\begin{figure}[t!]
\includegraphics[width=0.9\columnwidth]{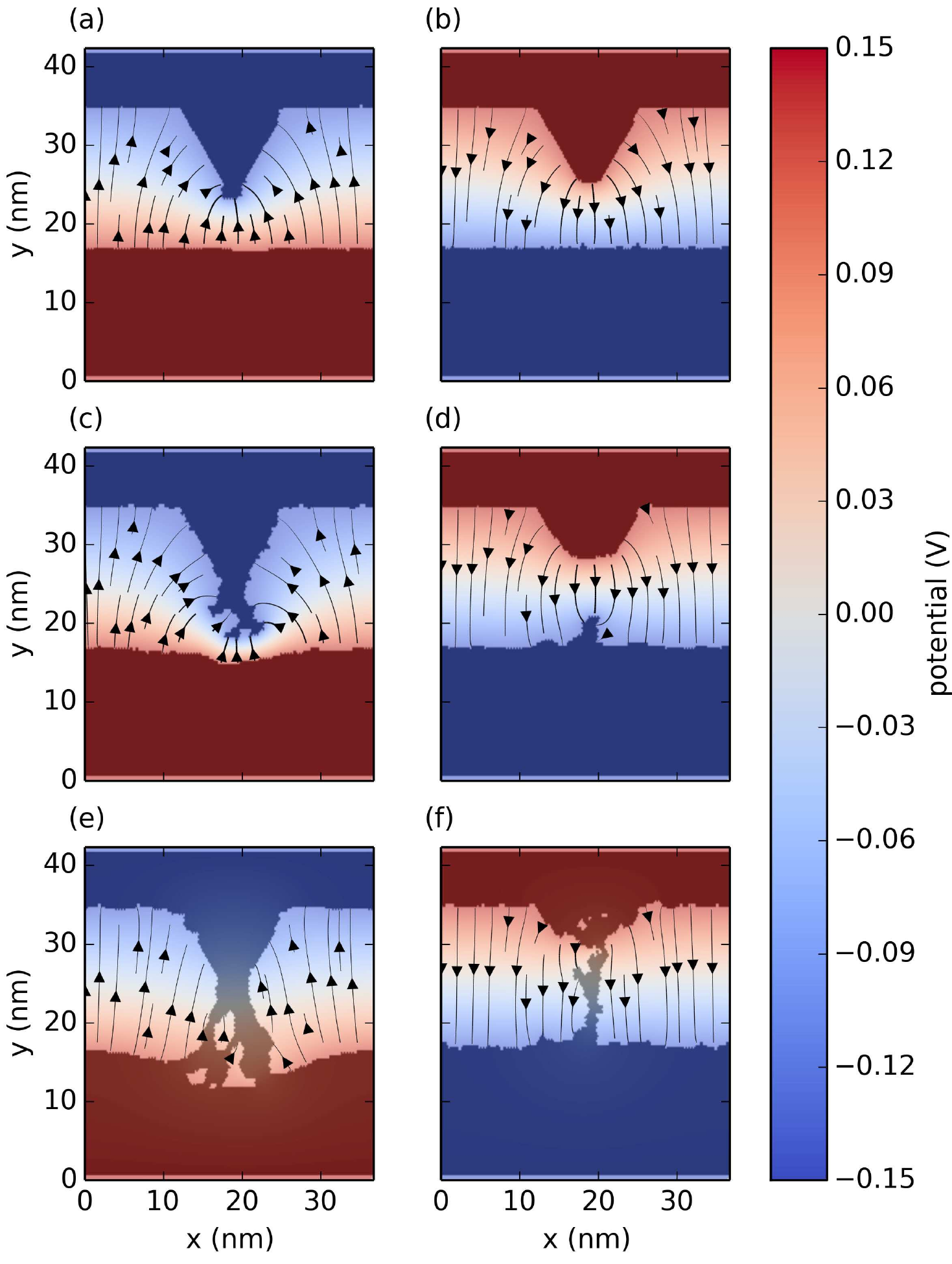}
\caption{Snapshots of initial silver filament formation within the Ag$_{2}$S layer at an asymmetric initial tip versus flat surface arrangement of the Ag
electrodes. The semi-transparent color map indicates the electrostatic potential and the stream lines visualize the electric field direction and magnitude
across the Ag$_{2}$S layer. The left and right panels show the junction's evolution for negative and positive tip potential $V_{\rm bias}$, respectively, at
identical initial geometries. The time evolution of the filamentary structure can be followed from the top to bottom panels. The complete structural evolution
can be seen in \href{https://youtu.be/W9CEvGQ-yew}{Animation 1}.} \label{T1.fig}
\end{figure}

\begin{figure}[t!]
\includegraphics[width=0.9\columnwidth]{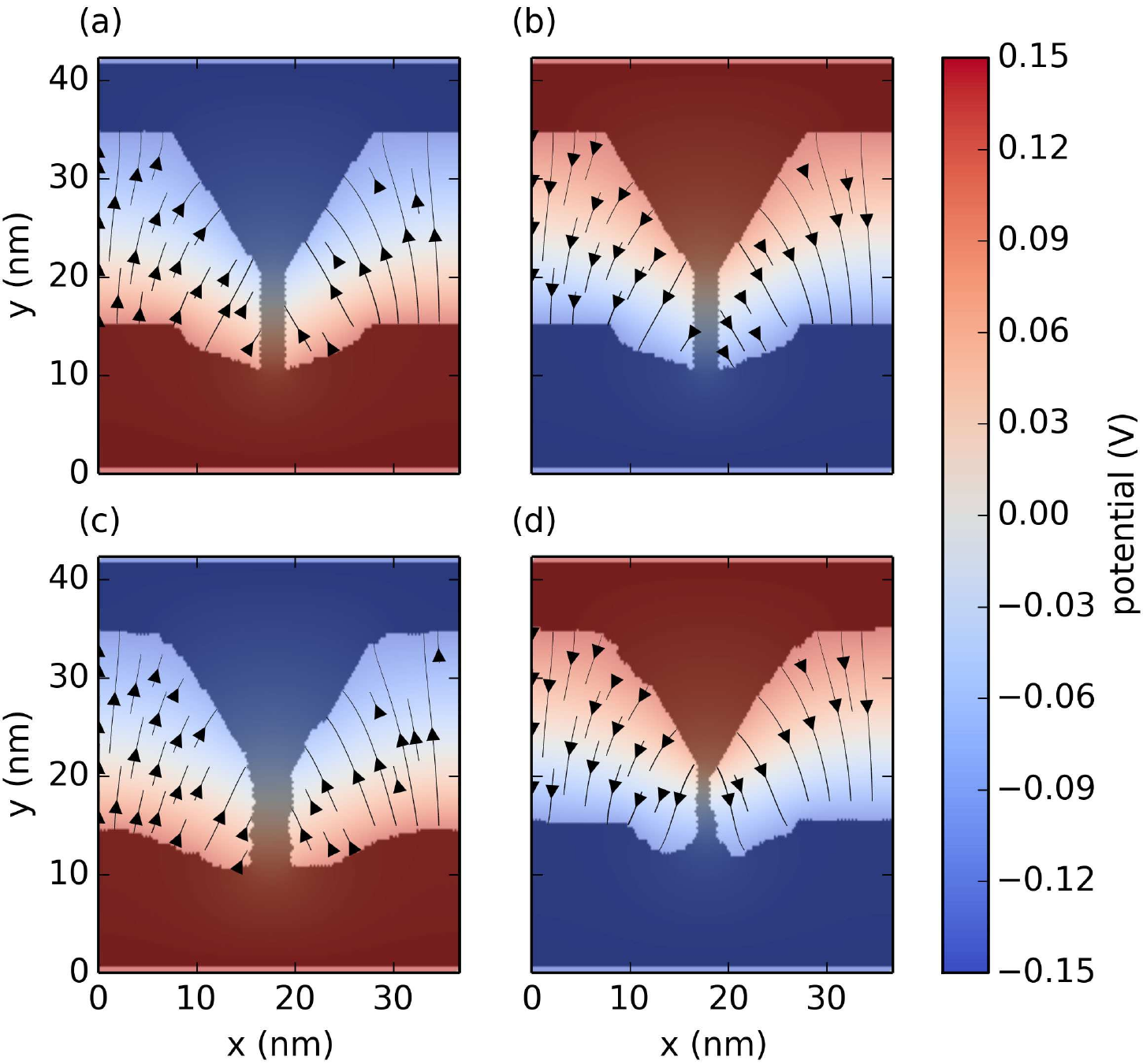}
\caption{(color online) Time evolution of a metallic silver junction bridging asymmetric Ag electrodes under opposite bias voltages. The semi-transparent color
map indicates the electrostatic potential and the stream lines visualize the electric field direction and magnitude across the Ag$_{2}$S layer. The top panels
show identical starting geometries. The bottom panels show the structure 1000 time steps later. The complete structural evolution is provided in
\href{https://youtu.be/DWqTSSCZ8GM}{Animation 2}.} \label{T2.fig}
\end{figure}

The simulations were carried out on a two-dimensional equilateral triangular lattice, where the lattice sites are either empty or occupied by a silver ion or
atom. The time development is performed either by moving some of the silver ions or atoms to their neighboring empty site or by simulating a redox reaction, in
which silver ions and atoms located at an electrode surface are exchanged. First the electrostatic potential is computed in each time step. This is followed by
the calculation of a transition probability for each possible change. Finally the changes are executed with the calculated probabilities. The transition
probabilities are computed as $\min\left(1,\frac{\Delta t}{\tau}e^{-\frac{\Delta E}{k_{B} T}}\right)$ where $\Delta E$ is the energy cost of the displacement
or the redox reaction, $1/\tau$ is the attempt rate and $\Delta t$ is the duration of the time step. The $\Delta E$ energy change depends on the participating
atom's or ion's interaction with its neighbors. In case of silver ions it also depends on the electrostatic potential. The ion-ion and ion-atom interactions
are parameterized close to room temperature in order to keep the silver ions sufficiently mobile. The atom-atom interaction is set to be strongly attractive
enabling the growth of stable metallic branches which resist to thermal diffusion. Further technical details of the simulations are available in the Appendix.
We emphasize that the two-dimensional aspect of the above model along with the assumption of a triangular lattice and the phenomenological transition
probabilities obviously cannot account for the rich variety of the microscopic details present in real Ag$_{2}$S nanojunctions. Yet, the reduced computational
requirements of such a simplified model allowed the analysis of several different parameter sets and boundary conditions and thus provided a deeper
understanding on the actual tendencies of electric field driven filament evolution at asymmetric electrode configurations.

Figure~\ref{T1.fig} shows a typical simulated evolution of the junction at negative and positive tip potentials (left and right panels in Fig.~\ref{T1.fig},
respectively). The structural development can be followed from the top to bottom panels. The initial asymmetrical arrangement, representing an STM tip - flat
surface setup, develops in time very differently at opposite bias polarities. Figures~\ref{T1.fig}(a) and \ref{T1.fig}(b) demonstrate that at the initial phase
of the filament formation the region of the most intensive structural changes are located at the apex of the tip, where the electric field is the highest. At a
negative tip potential the silver atoms are being deposited on the apex and a filament starts growing towards the bottom surface. Figure~\ref{T1.fig}(c)
illustrates a dendritic filament growth which is predominantly fueled by the oxidation of the silver atoms of the bottom electrode located under the filament.
At a reversed polarity the tip gradually loses its apex leading to an effective shrinking while opposite filament growth starting from the flat surface toward
the tip is taking place [Fig.~\ref{T1.fig}(d)]. Figures~\ref{T1.fig}(e) and \ref{T1.fig}(f) show the filament structures after establishing a metallic contact
between the electrodes. Figure~\ref{T1.fig}(e) demonstrates that the initial asymmetry of the contact is qualitatively preserved, while dendritic features in
the filament and a shallow dip in the planar electrode also appear. These qualitative features are in remarkable agreement with in-situ experimental studies on
the dynamics of nanoscale metallic inclusions in dielectrics \cite{Xu2010}, where due to the high redox rates and ion mobility \cite{Yang2013} in
stoichiometric Ag$_{2}$S, qualitatively similar filament growth was observed.

\begin{figure}[t!]
\includegraphics[width=\columnwidth]{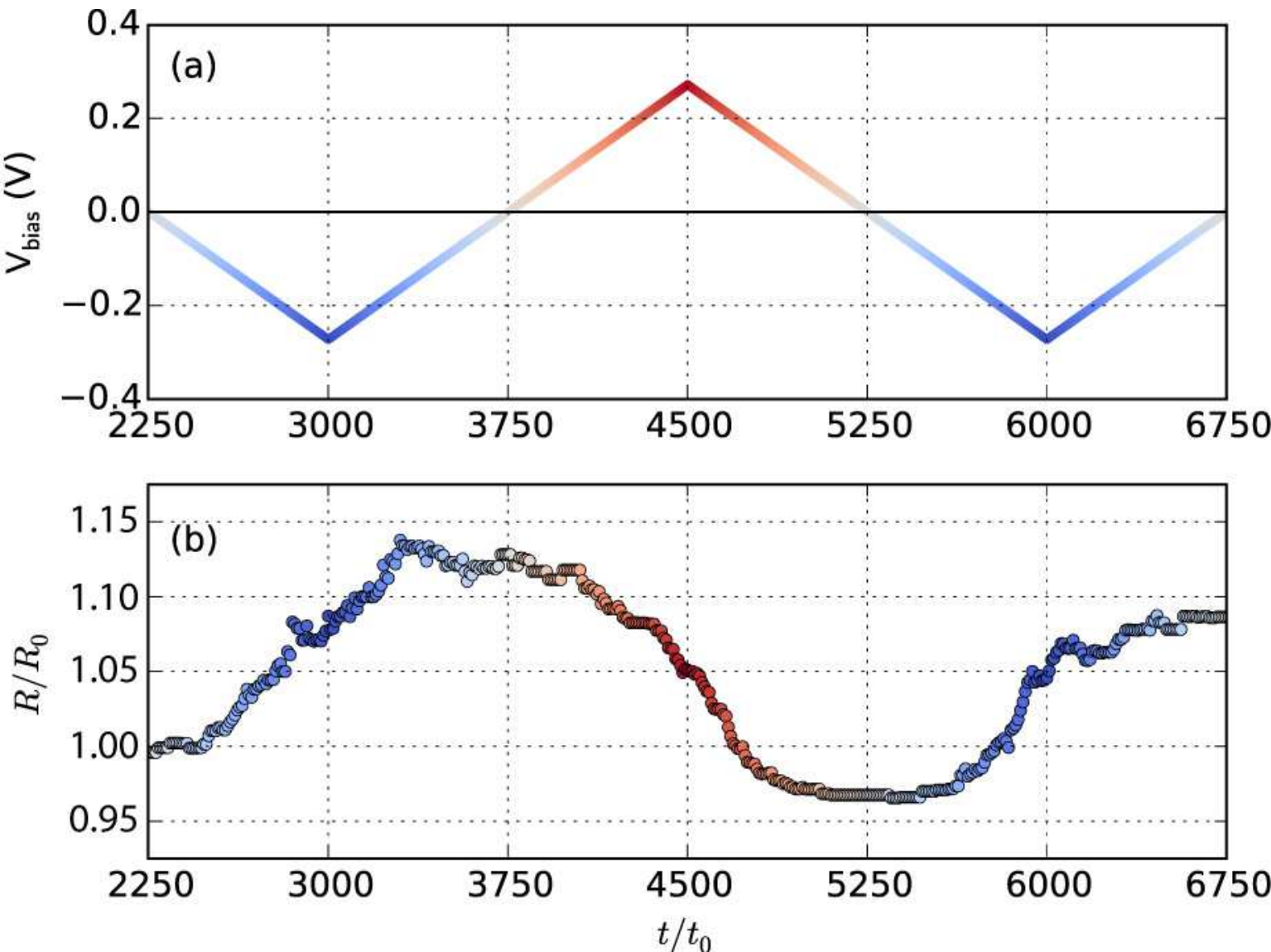}
\caption{Simulated resistive switching upon a time-dependent bias voltage. (a) Time dependence of the bias voltage. (b) The corresponding resistance across the
nanojunction. The complete structural evolution can be seen in \href{https://youtu.be/DWqTSSCZ8GM}{Animation 2}. $R_{0}$ is a normalization factor accounting
for the two-dimensional aspects of the simulation.} \label{T3.fig}
\end{figure}

Next we study the relation of the preserved asymmetry after complete filament formation to the observed, robustly uniform polarity of the resistive switchings
taking place between metallic OFF and ON states, corresponding to our experimentally investigated configuration [Fig.~\ref{stm.fig}]. We performed simulations
starting from simplified asymmetric filament boundaries representing an STM tip - flat surface setup as illustrated in Figs.~\ref{T2.fig}(a) and
\ref{T2.fig}(b). Figures~\ref{T2.fig}(c) and \ref{T2.fig}(d) demonstrate that under the influence of a negative (positive) tip bias the diameter of the
filament increases (reduces) by time, in full agreement with the observed polarities of the set/reset transitions, underlining the fundamental role of the
asymmetrical initial arrangement of the electrodes. The resulting electric field profiles reveal that the high field regions are located around the smallest
cross-section of the contact, so that resistive switching can occur due to the increased probabilities of redox steps in the vicinity of the filament.

The simulated resistive states upon repeated biasing cycles are shown in Fig.~\ref{T3.fig}. The two-terminal resistance is calculated from the bias voltage and
the current through the contact which is $I\sim-\int\nabla u\cdot d\mathbf{\hat{y}}$ where the integration is over the horizontal cross-section of the
two-dimensional plane. Due to its reduced dimensionality, our simple model is not expected to provide a quantitative agreement with the experimentally observed
resistance values nor to account for the closing of the hysteresis loops upon a biasing cycle. Nevertheless, calculating the two-terminal resistance by taking,
as a rude estimate, the bulk conductivity of silver and the narrowest (one-dimensional) cross-section of the junction into account, the basic qualitative
features of the hysteretic I-V traces can be reproduced.

\begin{figure}[t!]
\includegraphics[width=0.7\columnwidth]{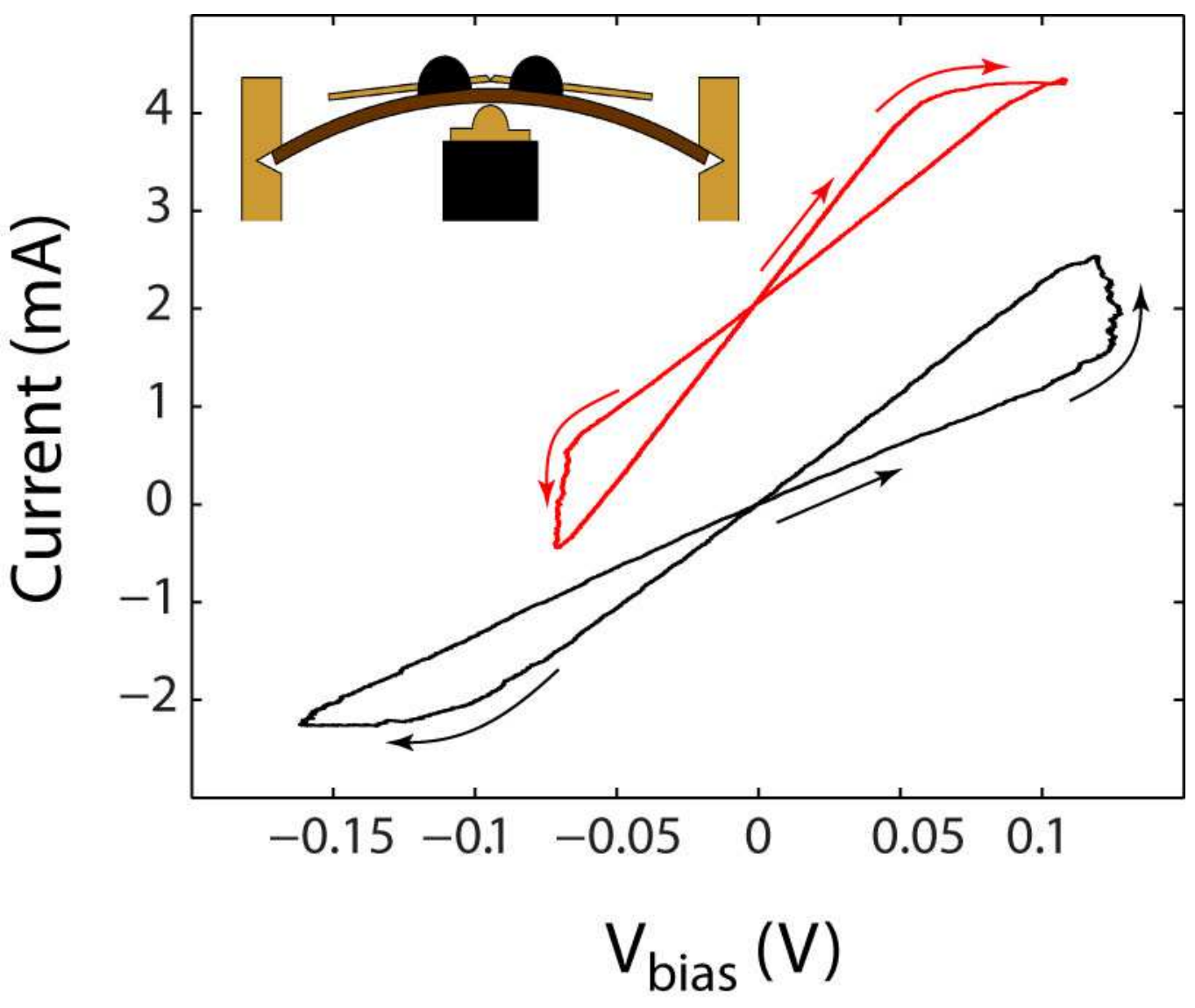}
\caption{(color online) Representative I-V traces recorded in selected, independent junctions established in an MCBJ arrangement by a 2.5~Hz triangular $V_{\rm
drive}$ voltage signal. The straight arrows indicate the random initial configurations measured after the creation of the junction at identical polarities. The
hysteresis loops exhibit a random, uniform distribution of either clockwise or anti-clockwise direction. The upper (red) curve is vertically shifted for
clarity. $R_{\rm OFF}$=77~$\Omega$, $R_{\rm ON}$=45~$\Omega$ and $R_{\rm S}$=150~$\Omega$ (black trace), $R_{\rm OFF}$=45~$\Omega$, $R_{\rm ON}$=25~$\Omega$
and $R_{\rm S}$=380~$\Omega$ (red trace). The inset illustrates the MCBJ setup.} \label{mcbj.fig}
\end{figure}

In order to further verify the dominant role of the geometrical asymmetry in the polarity of the resistive switchings, we also performed experiments on
sulfurized Ag-Ag junctions established in an MCBJ arrangement, as illustrated in the inset of Fig.~\ref{mcbj.fig}. We note that our previous experiments
\cite{C0NR00951B} utilizing the MCBJ technique for creating clean Ag-Ag atomic junctions at room temperature also reproduced the main features of the usually
observed resistive switching behavior which were attributed to electromigration taking place finite bias. However, in the absence of an ionic conductor layer
these individual characteristics were highly unstable in spite of the superior mechanical stability offered by the MCBJ technique \cite{Muller1992} over the
one of an off-feedback STM setup.

Unlike in the STM setup, where the initial asymmetry of the junction is largely predetermined by the different shapes of the thin film and sharp tip
electrodes, the controlled rupture of a uniform wire is expected to result in a randomly oriented local asymmetry at the apex of the nanojunction. In our
present experiments stable, hysteretic I-V traces exhibiting comparable, metallic ON and OFF state resistances but opposite switching polarities were obtained
after sulfurisation and controlled re-connection of the freshly created Ag-Ag junctions, as exemplified in Fig.~\ref{mcbj.fig}. The statistical analysis of 10
identically prepared samples revealed a 50\% probability of having a set/reset transition at a given voltage polarity independently from the applied bias
during re-connection, providing an excellent agreement with our qualitative scenario. When a stable sequence of I-V measurements was followed by a complete
rupture and re-connection of the electrodes, the polarity of the subsequent switchings were likely ($>$80\%) to be identical to those obtained previously,
indicating that up to a certain degree of re-establishment the asymmetry of the junction is mostly determined by the first rupture and is robust against
repeated mechanical reconfigurations.

\begin{figure}[t!]
\includegraphics[width=0.7\columnwidth]{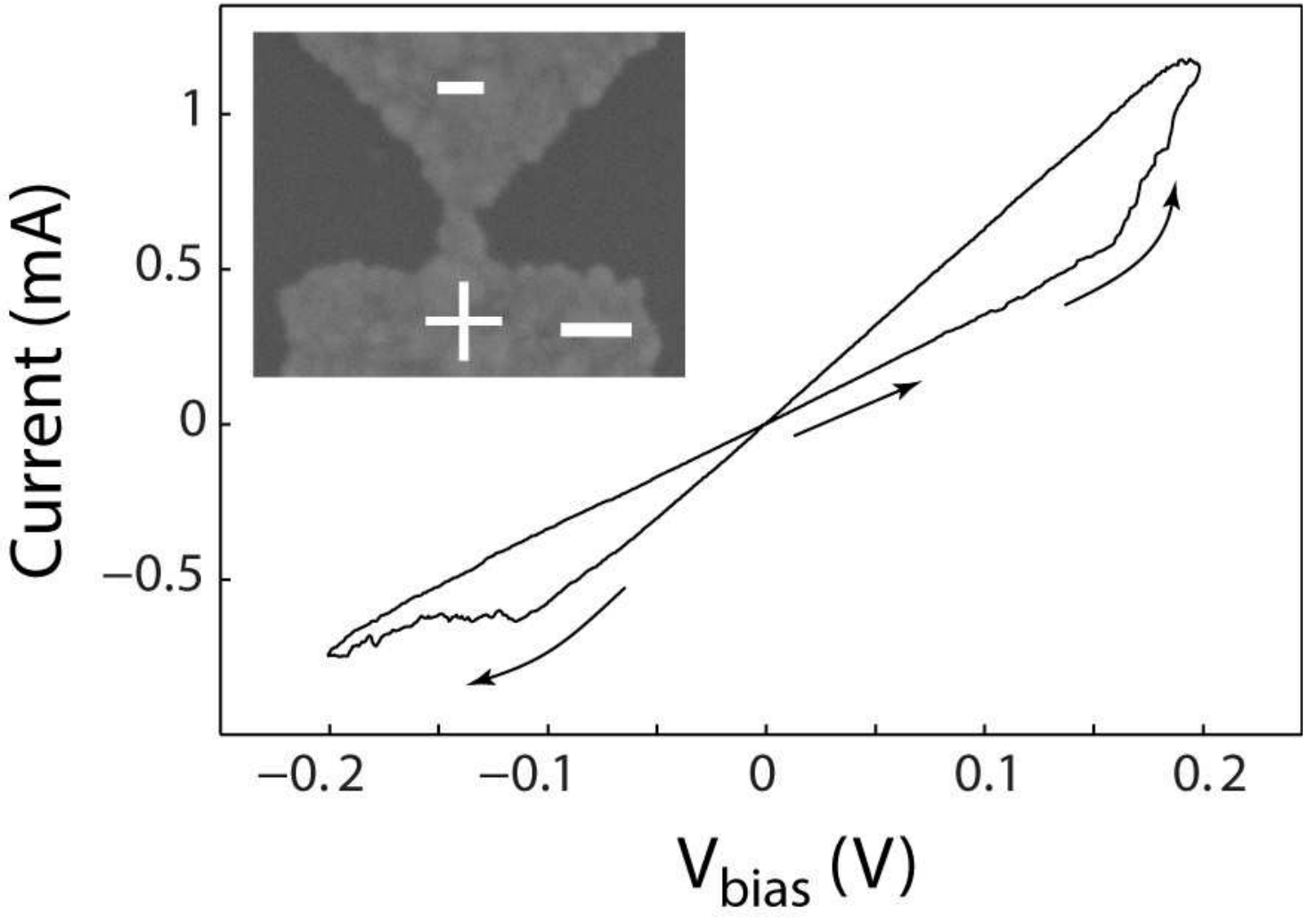}
\caption{Representative I-V trace recorded by a 2.5~Hz triangular $V_{\rm drive}$ voltage signal in a planar Ag-Ag$_{2}$S-Ag nanojunction created by
electromigration and subsequent sulfurisation of an all-Ag lithographic structure. $R_{\rm OFF}$=279~$\Omega$, $R_{\rm ON}$=155~$\Omega$ and $R_{\rm
S}$=5~$\Omega$. The inset shows the electron microscopy image of the device after performing the I-V measurements. The white scale bar in the lower right
corner indicates 200~nm. The convention of the bias voltage polarity is also shown.} \label{litho.fig}
\end{figure}

Based on the above findings we studied resistive switchings also in a series of prototype on-chip memory devices illustrated in the inset of
Fig.~\ref{litho.fig}. The structure mimicking the asymmetry of the STM arrangement was patterned by standard electron beam lithography on an amorphous, 140~nm
thick SiN$_{\rm x}$ substrate. The 100~nm wide and 45~nm thick, electron beam evaporated silver channel connecting the electrodes was further reduced in its
diameter by controlled electromigration in vacuum conditions \cite{Strachan2005,Esen2005,Nef2014}. During the electromigration process the junction was exposed
to a series of 0.5~ms long voltage pulses of increasing amplitude ranging from 10~mV to a maximum of 300~mV. The sample's resistance was monitored during the
100~ms dwell time between the pulses by acquiring low bias I-V traces. By repeating this method, the typical starting resistance of 50~$\Omega$ could be
increased to 250~$\Omega$. At this setpoint the bias was removed and due to a self-breaking mechanism \cite{Oneill2007,Prins2009} an approximately 1~nm wide
gap opened at the narrowest cross-section of the silver channel situated presumably close to the apex of the triangular electrode, as suggested by the inset of
Fig.~\ref{litho.fig}. This nanogap was then exposed to vaporized sulfur at 60$^{\circ}$C at ambient pressure followed by the re-connection of the electrodes by
triangular voltage signals up to 2~V amplitudes. An optimized, 3 minutes long sulfurisation time resulted stable, hysteretic I-V traces highly similar to those
obtained by the previously discussed techniques, also reflecting identical switching polarities as exemplified in Fig.~\ref{litho.fig}. This effect is
attributed to the formation of a few 10~nm long region of Ag$_{2}$S where filament formation and destruction can take place between the electrodes during
memory operations. While the endurance of these very first proof of concept devices were in the order of a few 100 cycles we believe that, owing to their
largely simplified structure and inherently high mechanical stability, the future optimization of the design and fabrication parameters is expected to enable
the reliable application of such memory architectures.

\begin{figure}[t!]
\includegraphics[width=\columnwidth]{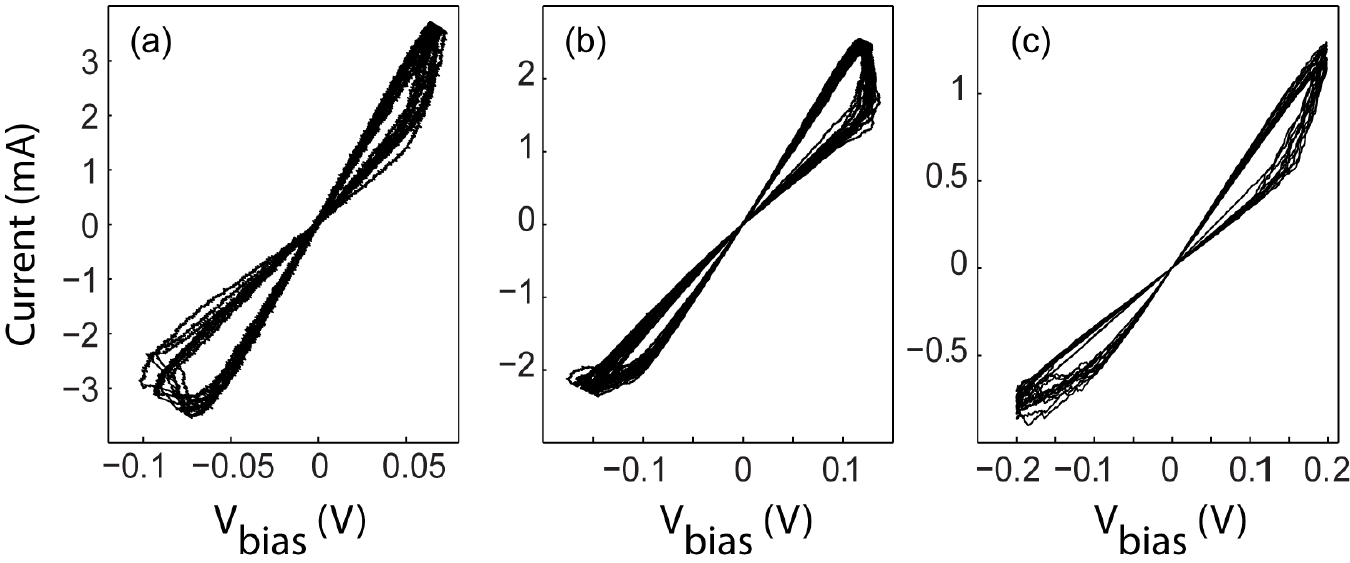}
\caption{20 consecutive I-V traces acquired on stable nanojunction configurations established in an STM setup (a), MCBJ setup (b) and in a nanolithographic
sample (c). $R_{\rm OFF}$=32.6$\pm$3.3~$\Omega$, $R_{\rm ON}$=18.2$\pm$1.6~$\Omega$ (a), $R_{\rm OFF}$=77$\pm$5~$\Omega$, $R_{\rm ON}$=45$\pm$1.8~$\Omega$ (b)
and $R_{\rm OFF}$=279$\pm$15~$\Omega$, $R_{\rm ON}$=155$\pm$5~$\Omega$ (c).} \label{stab.fig}
\end{figure}

The reproducibility of the measured I-V traces at stable junction configurations are illustrated in Figs.~\ref{stab.fig}(a), \ref{stab.fig}(b) and
\ref{stab.fig}(c) for STM, MCBJ and lithographically designed on-chip structures, respectively. The evaluated average $R_{\rm ON}$ and $R_{\rm OFF}$ values
deduced from the curves shown in Fig.~\ref{stab.fig}(a) and their $\sim$10\% relative standard deviation comply with those obtained in similarly established
PtIr-Ag$_{2}$S-Ag nanojunctions \cite{Gubicza2015temp} and are limited by the long-term mechanical stability of our off-feedback STM setup \cite{Gubicza2015}.
The superior mechanical stability of the MCBJ setup over the one of the off-feedback STM is reflected in the reduced relative standard deviations of the
$R_{\rm OFF}$ and $R_{\rm ON}$ values in Fig.~\ref{stab.fig}(b). While the STM and MCBJ setups were installed by implementing advanced isolation techniques
against mechanical vibrations, the I-V measurements of the lithographic structures were carried out in a mechanically undamped vacuum chamber which was
installed directly at the inlet of a turbomolecular pump. Yet, the lowest dispersion of the $R_{\rm OFF}$ and $R_{\rm ON}$ values was achieved in this setup
[Fig.~\ref{stab.fig}(c)], demonstrating the inherently high mechanical stability and robustness of our simplified on-chip design.

In conclusion, we investigated stable resistive switchings in Ag-Ag$_{2}$S-Ag nanojunctions lacking the conventionally employed inert electrode. Our
experiments performed in the STM and MCBJ arrangements demonstrated that the polarity of the set/reset transitions are exclusively determined by the
inhomogeneity of the local electric field, arising from the geometrical asymmetry present at the apex of the junction. Numerical simulations taking activated
ion migration and redox reactions into account successfully reproduced the observed switching behavior also in the so-far-less-widely-investigated metallic
regime. The simulations also reveal that the atomic re-arrangements responsible for the observed resistive switchings only involve a small amount of silver
ions situated in the vicinity of the junction's narrowest cross-section providing a key ingredient to ultrafast memory operation \cite{GeresdiAndreev}. Our
proof of principle experiments demonstrate the merits of lithographically designed Ag-Ag$_{2}$S-Ag nanostructures as fast and highly integrable memory cells.
Additionally, by further optimization of the nanometer-scale, planar on-chip design, an inherently high mechanical stability is envisioned, whereas the
utilization of all-Ag electrodes makes the lithographical fabrication procedure uncomplicated.

This work was supported by the Hungarian Research Funds OTKA K105735, K112918, the EC FP7 ``CoSpinNano'' (project No. 293797) and ITN ``MOLESCO'' (project No.
606728) and by the UK EPSRC grants EP/K001507/1, EP/J014753/1 and EP/H035818/1. Useful discussions with Z. Balogh, B. F\"ul\"op and M. Calame are acknowledged.
The experimental data is held by the Department of Physics, Budapest University of Technology and Economics. The underlying data for the theory analysis is
available in the Appendix.

\appendix*
\section{Appendix: The numerical simulations}

In order to qualitatively investigate the silver filament growth in Ag-Ag$_{2}$S-Ag structures we performed atomic scale two-dimensional lattice based
simulations. The simulation is implemented on an equilateral triangular lattice where a site can represent an empty site, a silver atom or a mobile silver ion.
The presence of sulphide ions is taken into account as a screening medium reducing the range of the silver ion interactions to nearest neighbor. The
microscopic development is driven by room temperature ionic diffusion and redox processes, the latter taking place at the electrode surfaces. Similar filament
growth simulations have been done before in Ref.~\citenum{zhang2014quantized} utilizing different theoretical models.

A typical equilateral triangular lattice used in the simulations can be recognized in Fig.~\ref{S1.fig} where the black dots denote silver atoms, the red dots
represent silver ions. The lattice exhibits periodic boundary conditions along the horizontal direction while the vertical boundaries are terminated by two
layers of silver atoms set to $V$ and zero potentials on the top and bottom, respectively. The lattice constant is set to $a$=3.85~${\AA}$ which approximately
corresponds to one real atom per site. The electrostatic potential on each site is computed by solving the Poisson's equation $\nabla(\epsilon_{r}\nabla
u)=-\rho/\epsilon_{0}$ on the lattice. The charges of the silver ions are considered to be screened, therefore they are excluded from the electrostatic
calculation. $\epsilon_{r}=1$ is set to zero outside, whereas inside the silver $\epsilon_{r}=1-i1.25\times10^{5}$ is applied. The surface charge density is
computed from the potential as $\triangle u=-\rho/\epsilon_{0}$.

The time development is performed either by moving some of the silver ions or atoms to their neighboring empty site or by simulating a redox reaction, in which
silver ions and atoms located at an electrode surface are exchanged. First the electrostatic potential is computed in each time step. This is followed by the
calculation of a transition probability for each possible change. Finally the changes are executed with the calculated probabilities. The transition
probability of a silver ion at site $k_{+}$ moving to its neighboring empty site $k_{\circ}$ is computed as \cite{domb1995statistical}
\[
w_{k_{+},k_{\circ}}^{\mathrm{diff}}=\min\left(1,\frac{\Delta t}{\tau_{+}}e^{-\frac{\Delta E_{k_{+},k_{\circ}}^{\mathrm{diff}}}{k_{B}T}}\right),
\]
where $\Delta E_{k_{+},k_{\circ}}^{\mathrm{diff}}$ is the energy change of the move, $1/\tau_{+}$ is the attempt frequency of the silver ion to jump and
$\Delta t$ is the duration of the time steps. The $w_{k_{\bullet},k_{\circ}}^{\mathrm{diff}}$ transition probability of a silver atom jumping from site
$k_{\bullet}$ to its neighboring empty site $k_{\circ}$ is computed similarly using $\Delta E_{k_{\bullet},k_{\circ}}^{\mathrm{diff}}$ and $1/\tau_{\bullet}$.
The transition probability for a redox step, where a surface silver atom on site $k_{\bullet}$ is oxidized and a surface silver ion on site $k_{+}$is reduced,
is computed as
\[
w_{k_{\bullet},k_{+}}^{\rm redox}=\min\left(1,\frac{\Delta t}{\tau_{\rm redox}}e^{-\frac{\Delta E_{k_{\bullet}}^{\mathrm{ox}}+\Delta
E_{k_{+}}^{\mathrm{red}}}{k_{B}T}}\right),
\]
where $\Delta E_{k_{\bullet}}^{\mathrm{ox}}+\Delta E_{k_{+}}^{\mathrm{red}}$ is the energy change due to the oxidation and reduction on site $k_{\bullet}$ and
site $k_{+}$, respectively, and $1/\tau_{\rm redox}$ is the redox reaction rate. The $\Delta t$ duration of the time steps is chosen such that the typical
transition probability is much smaller than 1. The energy changes are calculated as
\[
\Delta E_{k_{+},k_{\circ}}^{\mathrm{diff}}=\tilde{\mu}_{k_{\circ}}(Ag^{+})-\tilde{\mu}_{k_{+}}(Ag^{+})
\]
\[
\Delta E_{k_{\bullet},k_{\circ}}^{\mathrm{diff}}=\tilde{\mu}_{k_{\circ}}(Ag)-\tilde{\mu}_{k_{\bullet}}(Ag)
\]
\[
\Delta E_{k_{\bullet}}^{\mathrm{ox}}=\tilde{\mu}_{k_{\bullet}}(Ag^{+})-\tilde{\mu}_{k_{\bullet}}(Ag)+\tilde{\mu}_{k_{\bullet}'}^{e}
\]
\[
\Delta E_{k_{+}}^{\mathrm{red}}=\tilde{\mu}_{k_{+}}(Ag)-\tilde{\mu}_{k_{+}}(Ag^{+})-\tilde{\mu}_{k_{+}'}^{e}
\]
where the electrochemical potential of a silver ion at site $k$ is
\[
\tilde{\mu}_{k}(Ag^{+})=\gamma_{++}n_{k}^{+}+\gamma_{+\bullet}n_{k}^{\bullet}+|e|u_{k},
\]
where $n_{k}^{+}$and $n_{k}^{\bullet}$ are the numbers of silver ion and atom neighbors of site $k$, respectively, $\gamma_{++}$ and $\gamma_{+\bullet}$ are
the interaction energies, $|e|$ is the charge of the silver ion and $u_{k}$ is the potential at site $k$. The electrochemical potential for the silver atom is
\[
\tilde{\mu}_{k}(Ag)=\gamma_{+\bullet}n_{k}^{+}+\gamma_{\bullet\bullet}n_{k}^{\bullet},
\]
where the $\gamma_{\bullet\bullet}$ is the interaction energy between silver atoms. The sites $k_{\bullet}'$ and $k_{+}'$ denote neighboring silver sites to
$k_{\bullet}$ and $k_{+}$, respectively. They are chosen to provide the maximum probability for the given redox reaction. The electron's electrochemical
potential on a surface site is calculated as \cite{bernard2003mean}
\[
\tilde{\mu}_{k}^{e}=E_{F}-|e|u_{k}-\frac{\rho_{k}/|e|}{g(E_{F})},
\]
where $\rho_{k}$ is the charge density at the surface site $k$ and $g(E_{F})$ is the surface density of states in silver. The structural evolution of the
contact is determined by the locations of the redox processes. The above formulae of the chemical potential imply that the typical locations of the redox
reactions are determined by the average numbers of neighbors i.e., the ion concentration, and by the surface charge density. The numbers of neighbors are
weighted with the interaction energies, therefore this contribution is coupled to the energetics of the diffusion process, whereas the surface charge density
terms depend on the metal properties.

\renewcommand\thefigure{A\arabic{figure}}
\setcounter{figure}{0}

\begin{figure}[t!]
\includegraphics[width=0.97\columnwidth]{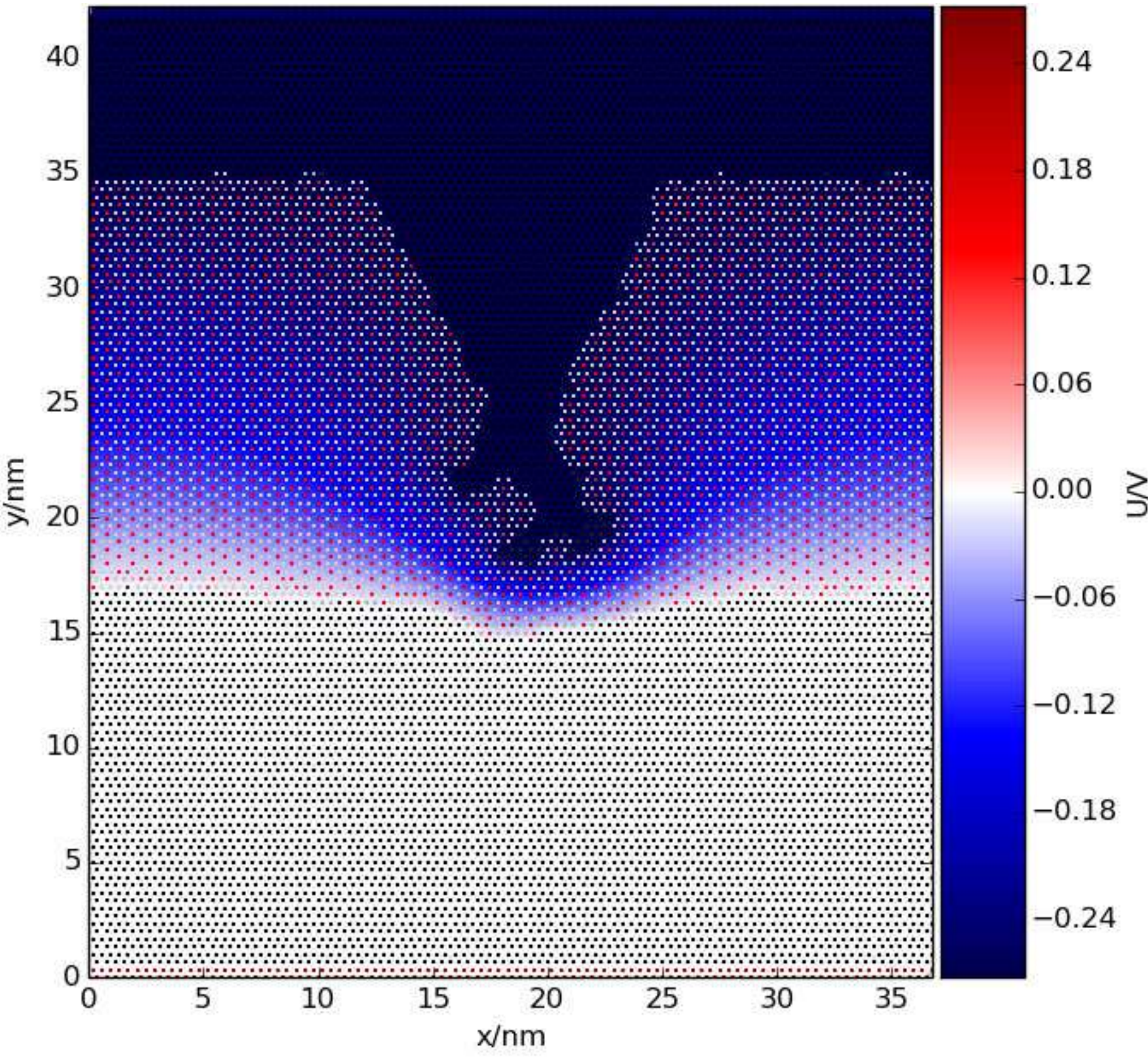}
\caption{Snapshot of initial silver filament formation within the Ag$_{2}$S layer at an asymmetric tip versus flat surface arrangement of the Ag electrodes.
The semi-transparent color map indicates the electrostatic potential. The black (red) dots represent Ag atoms (Ag$^{+}$ ions) The grey dots denote empty
sites.} \label{S1.fig}
\end{figure}

\begin{figure}[t!]
\includegraphics[width=\columnwidth]{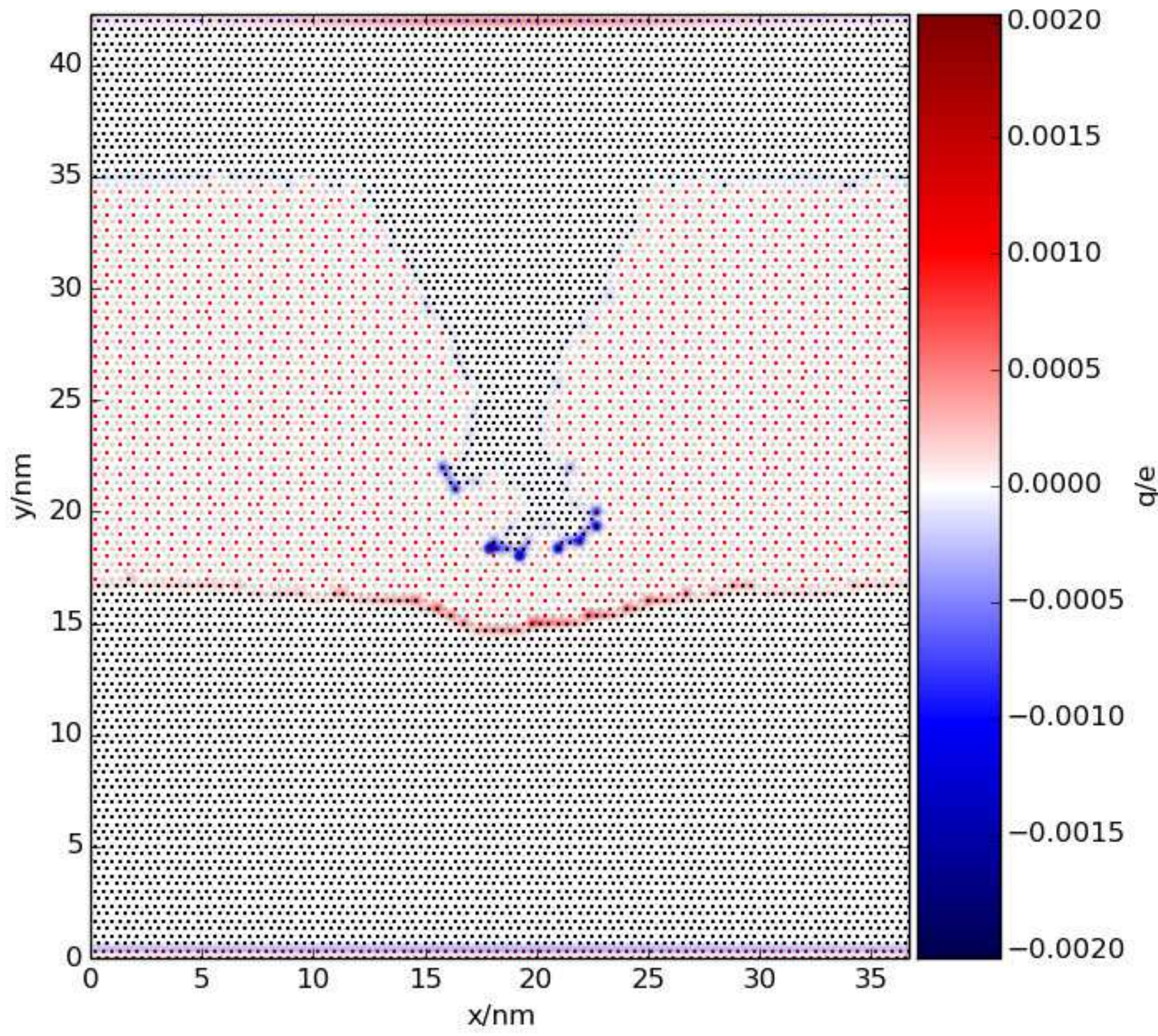}
\caption{Color map of the surface charge density in the arrangement of Fig.~\ref{S1.fig}. The black (red) dots represent Ag atoms (Ag$^{+}$ ions). The grey
dots denote empty sites.} \label{S2.fig}
\end{figure}

The typical parameter values utilized in the simulations were $\tau_{+}/\tau_{\rm redox}$=0.001, $\tau_{\bullet}/\tau_{\rm redox}$=0.01,
$\gamma_{++}/k_{B}T$=3, $\gamma_{+\bullet}/k_{B}T$=2, $\gamma_{\bullet\bullet}/k_{B}T$=-20, $k_{B}Tg(E_{F})a^{3}\sqrt{3}=3\times10^{-5}$, $eV/k_{B}T$=10, and
the concentrations of mobile silver ions, $c\approx1/3$. Adjusting the parameter values two extreme scenarios may take place: the initial structure either
explodes or no significant change occurs during the simulated timescale. Trying different parameter sets to avoid such extreme outcomes, we found that
resistive switching is a robust phenomenon in the simulated memristor structures which does not require fine tuning of the parameters. Increasing the voltage
bias or reducing the $\gamma_{\bullet\bullet}/k_{B}T$ strength of the silver-silver interactions can lead to more dendritic evolvement and also the structure
more likely explodes to a random mixture of atoms and ions. The $k_{B}Tg(E_{F})a^{3}\sqrt{3}$ parameter provides a means to increase the probability of redox
reactions without tuning the diffusion parameters.

A snapshot of the electrostatic potential and surface charge density maps corresponding to a dendritic filament formation are displayed along with the
underlying lattice in high resolution in Figs.~\ref{S1.fig} and \ref{S2.fig}, respectively. Two animation files are also added to the present paper.
\href{https://youtu.be/W9CEvGQ-yew}{Animation 1} shows the structural evolution during silver filament formation in an asymmetrical tip - flat surface
electrode arrangement, as explained via the snapshots displayed in Fig.~\ref{T1.fig}. \href{https://youtu.be/DWqTSSCZ8GM}{Animation 2} shows the modulation in
the width of a completed metallic filament having asymmetrical boundaries (as seen in Fig.~\ref{T2.fig}) under bias voltages of alternating sign.

%\href{https://youtu.be/W9CEvGQ-yew}{Animation 1}
%\href{https://youtu.be/DWqTSSCZ8GM}{Animation 2}

%\bibliographystyle{apsrev}
%\bibliography{References}

\end{document}